\newif\ifeprint\eprinttrue

\ifeprint

\documentclass[rmp,twocolumn,amsmath,showkeys,letterpaper]{revtex4}
\makeatletter\def\raggedcolumn@skip{\vskip\z@\@plus.0001fil\relax}\makeatother
\def\tbl#1#2{\caption{#1}#2}
\def\TITLE{Sampling exactly from the normal distribution}

\else

\documentclass[prodmode,acmtoms]{acmsmall}
\def\citet{\citeN}
\def\citep{\cite}
\usepackage{amsmath}
\usepackage{amssymb}
\def\TITLE{Sampling Exactly from the Normal Distribution}

\fi

\newcommand{\slug}{\hbox{\kern1.5pt\vrule
width2.5pt height7pt depth0.5pt\kern1.5pt}}
\def\xskip{\hskip 7pt plus 3pt minus 4pt}
\newdimen\algindent
\newif\ifitempar \itempartrue 
\def\algindentset#1{\setbox0\hbox{{\bf #1.\kern.25em}}\algindent=\wd0\relax}
\def\algbegin #1 #2{\algindentset{#21}\alg #1 #2} 
\def\aalgbegin #1 #2{\algindentset{#211}\alg #1 #2} 
\def\alg#1(#2). {\medbreak 
  \noindent{\bf#1}({\it#2\/}).\xskip\ignorespaces}
\def\altalgbegin #1 {\algindentset{#11}\altalg}
\def\altalg(#1) {\medbreak 
  \noindent{\it#1\/}\xskip\ignorespaces}
\def\algstep#1.{\par\ifitempar\smallskip\noindent\else\itempartrue
  \hskip-\parindent\fi
  \hbox to\algindent{\bf\hfil #1.\kern.25em}%
  \hangindent=\algindent\hangafter=1\ignorespaces}
\def\algend{\mbox{\quad\slug}\medbreak\noindent}
\def\altalgend{\medbreak\noindent}

\usepackage{ifpdf}
\usepackage{times}
\usepackage[bookmarks=false]{hyperref}
\usepackage{breakurl}
\usepackage{listings}
\usepackage{graphicx}

\hypersetup{pdftitle={\TITLE},pdfauthor={C. F. F. Karney}}

\def\d{\mathrm d}
\def\abs#1{\left|#1\right|}
\def\ave#1{\left<#1\right>}
\def\floor#1{\lfloor#1\rfloor}
\def\ceil#1{\lceil#1\rceil}
\newcount\minutes \newcount\hours
\hours=\time \divide\hours by 60 \multiply\hours by 60
\minutes=\time \advance\minutes by -\hours \divide\hours by 60
\def\twodigit#1{\ifnum#1<10 0\fi\the#1}

\date{March 25, 2013; revised \today}

\def\DATE{\monthWord\month\ \the\day, \the\year}

\ifpdf\def\urlalt#1#2{\burlalt{#2}{#1}}\else\let\urlalt=\burlalt\fi
\def\doi#1{\urlalt{http://dx.doi.org/#1}{doi:#1}}

\begin{document}
\title{\TITLE}

\ifeprint

\author{Charles F. F. Karney}
\email{charles.karney@sri.com}
\affiliation{\href{http://www.sri.com}{SRI International},
201 Washington Rd, Princeton, NJ 08543-5300, USA}

\else

\markboth{C. F. F. Karney}{\TITLE}
\author{CHARLES F. F. KARNEY
\affil{SRI International}}
\acmformat{Charles F. F. Karney.  2014. \TITLE}

\begin{bottomstuff}
Author's address: Charles Karney, SRI International,
201 Washington Rd, Princeton, NJ 08543-5300.
Electronic address:
\href{mailto:charles.karney@sri.com}{charles.karney@sri.com}
\end{bottomstuff}

\fi


\begin{abstract}
An algorithm for sampling exactly from the normal distribution is given.
The algorithm reads some number of uniformly distributed random digits
in a given base and generates an initial portion of the representation
of a normal deviate in the same base.  Thereafter, uniform random digits
are copied directly into the representation of the normal deviate.
Thus, in contrast to existing methods, it is possible to generate normal
deviates exactly rounded to any precision with a mean cost that scales
linearly in the precision.  The method performs no extended precision
arithmetic, calls no transcendental functions, and, indeed, uses no
floating point arithmetic whatsoever; it uses only simple integer
operations.  It can easily be adapted to sample exactly from the
discrete normal distribution whose parameters are rational numbers.

\end{abstract}

\maketitle

\section{Introduction}\label{intro}

Random variables with a normal density,
\[
\phi(x) = \frac{\exp(-x^2/2)}{\sqrt{2\pi}},
\]
are widely used in Monte Carlo simulations.  Over the past sixty years,
scores of algorithms for generating such normal deviates have been
published \citep{thomas07}.  In this paper, I give another algorithm,
Algorithm N, with the distinguishing feature that, given a source of
uniformly distributed random digits in some base $b$, it generates {\it
exact} normal deviates.  In order to make the meaning of ``exact''
precise, consider Table \ref{example} which shows the operation of the
algorithm using $b=10$.

The first column shows the random digits used by the algorithm, which,
in this example, are taken from successive lines of the table of random
digits produced by the \citet{rand55}, beginning at line 9077.
Referring to the first line of this table, the algorithm completes after
reading the 7 random decimal digits, $9148686$, from the source and
produces $+1.6$ as the initial portion of the decimal representation of
a normal deviate.  At this point, random digits can be copied directly
from the input to the output, indicated by the ellipses ($\ldots$) in
the table; thus $+1.6\ldots$ represents a uniform random sample in the
range $(1.6, 1.7)$.  The next digits of the random sequence are
$685171\ldots$, allowing the normal deviate to be exactly rounded to 6
decimal digits as $1.668517$.  I call the intermediate result, e.g.,
$+1.6\ldots$, a ``u-rand''.  This can be thought of as a partially
sampled uniform deviate.  However, in conjunction which a source of
random digits it is better to think of it as a compact representation of
an arbitrary precision random deviate.  (The results in
Table \ref{example} are not ``typical,'' because the starting line in
the table of random digits was specifically chosen to limit the number
of random digits used.)
\begin{table}[tb]
\tbl{\label{example}Sample input and output for Algorithm N with $b = 10$.
The input consists of uniformly distributed random decimal digits.  The
algorithm reads the random digits before the vertical bar and produces a
normal deviate as a u-rand (given in the second column), which is an
initial portion of the decimal representation of the normal deviate.
Thereafter random digits are copied directly from the input (the digits
after the vertical bar) into the decimal fraction of the u-rand.  The
third column shows the result of adding enough digits to allow the
deviate to be rounded to 6 decimal digits; the parenthetical sign
indicates whether the magnitude of true deviate is greater $(+)$ or
smaller $(-)$ than the rounded result.}{%
\setlength{\tabcolsep}{6pt}
\begin{tabular}{lll}
\hline\hline\noalign{\smallskip}
input & \multicolumn{1}{l}{u-rand} & \multicolumn{1}{c}{rounded\quad\null}
\\\hline\noalign{\smallskip}
$9148686\mid685171\ldots$ & $+1.6\ldots$ & $+1.668517(+)$\\
$2708\mid5545979\ldots$ & $+0\ldots$ & $+0.554598(-)$\\
$501446297\mid43871\ldots$ & $+1.42\ldots$ & $+1.424387(+)$\\
$065130319777860\mid96289\ldots$ & $-0.76\ldots$ & $-0.769629(-)$\\
$2736\mid0659086\ldots$ & $+0\ldots$ & $+0.065909(-)$\\
\hline\hline
\end{tabular}
}
\end{table}

It's clear from this example that the method can be used to generate
deviates that satisfy the conditions of ``ideal
approximation'' \citep{monahan85}, namely that the algorithm is
equivalent to sampling a real number from the normal distribution and
rounding it to the nearest representable floating point number.
Furthermore, for applications requiring high precision normal deviates,
the new algorithm offers perfect scaling: there's an amortized constant
cost to producing the initial portion of the normal deviate; but,
thereafter, the digits can be added to the result at a rate limited only
by the cost of producing and copying the random digits.  Other sampling
methods are frequently referred to as ``exact,'' for example the polar
method \citep{box58} and the ratio method \citep{kinderman77}; but these
are merely ``accurate to round off'' which, in practice, means only that
the accuracy is commensurate with the precision of the floating point
number system.  It's possible to convert such algorithms to obtain
correctly rounded deviates; but this inevitably involves the use of
extended precision arithmetic.  I will show, in Sec.~\ref{implement},
that Algorithm N performs substantially better.

It's not immediately obvious that such an algorithm for exact sampling
is possible.  However, in the early years of the era of modern
computing, \citet{vonneumann51} presented a remarkably simple algorithm
for sampling from the exponential density, in which ``the machine has in
effect computed a logarithm by performing only discriminations on the
relative magnitude of numbers in $(0,1)$.'' \citet{knuth76} showed that
the algorithm can easily be adapted to generate exponential deviates
which are exact; and the resulting method was extensively analyzed
by \citet{flajolet86}.  Several authors have generalized von Neumann's
algorithm \citep{forsythe72,ahrens73,brent74,monahan79}.  However, these
efforts entail using ordinary floating point arithmetic and thus the
methods do not generate exact deviates.

In this paper, I show that von Neumann's algorithm can be extended to
sample exactly from the unit normal and discrete normal distributions.
Although the resulting algorithms are unlikely to displace existing
methods for most applications, it provides a nearly optimal method for
generating normal deviates at arbitrary precision.  In addition, the
ability to sample exactly from the discrete normal has applications to
cryptography because the security of cryptographic systems requires that
any random sampling be very accurate.  Finally, the method is of
theoretical interest as an example of an algorithm where exact
transcendental results can be achieved with simple integer arithmetic.

Implementations of the algorithms given in this paper are available in
ExRandom, a small ``header only'' library for C++11, available
at \href{http://exrandom.sf.net/} {\tt http://exrandom.sf.net/}, and in
version 3.2.0 of \citet{MPFR}.

\section{von Neumann's algorithm}\label{vonN}

I begin by reviewing von Neummann's algorithm, because this is the basis
of the method for sampling from the normal distribution.

\algbegin Algorithm V (von Neumann).  Samples $E$ from the exponential
distribution $e^{-x}$ for $x > 0$.
\algstep V1. [Initialize rejection count.] Set $l\leftarrow 0$.
\algstep V2. [Sample fraction.]  Set $x \leftarrow U$, where $U$
is a uniform deviate $U \in (0,1)$.
\algstep V3. [Generate a run.]  Sample uniform
deviates $U_1, U_2, \ldots$ and determine the maximum value $n \ge 0$
such that $x > U_1 > U_2 > \ldots > U_n$.
\algstep V4. [Test length of run.]  If $n$ is odd set,
$l \leftarrow l + 1$ and go to step V2.
\algstep V5. [Combine integer and fraction.]  Otherwise ($n$ is even),
set $y \leftarrow l + x$.
\algstep V6. [Return result.] Set $E \leftarrow y$.%
\algend
(Because the algorithm generates continuous random deviates, there's no
distinction between the inequalities $x \ge 0$ and $x > 0$ or the
intervals $(0,1)$ and $[0,1]$.)  According to von Neumann, this
algorithm was suggested by the game of Black Jack and this connection is
made plain in the slightly different formulation given
in \citet[\S26.8.6.c(2)]{abramowitz65}.

The crucial step of the algorithm is V3, which is discussed in some
detail by \citet[\S3.4.1.C(3)]{Knuth98}.  The probability that
$U_1, \ldots, U_n$ are all less than $x$ is $x^n$ (provided that $x \in
[0,1]$).  The probability that, in addition, they are in descending
order (one of the possible $n!$ permutations) is $x^n/n!$.  For the
condition to hold for a sequence of $n+1$ numbers, it must hold for the
first $n$ of them; therefore the probability that the length of the
longest decreasing sequence is $n$ is $x^n/n! - x^{n+1}/(n+1)!$.  For a
given $x$, the probability that $n$ is even is
\[
1 - x + \frac{x^2}{2!} - \frac{x^3}{3!} + \ldots = e^{-x},
\]
while the probability that $n$ is odd (averaged over $x$) is $1
- \int_0^1 e^{-x} \,\d x = e^{-1}$.  Thus the probability that the
algorithm terminates with a particular value of $l$ and $x$ is
$\exp\bigl(-(l + x)\bigr)$ as required.

On average, this algorithm requires $e^2/(e-1) \approx 4.30$ uniform
deviates; in effect, the algorithm sums all the terms in the Taylor
series for $e^{-x}$ in a finite mean time.  Conventionally, $U$ would be
sampled from the subset of reals which are representable as double
precision numbers; in this case, the results would be only approximately
equivalent to sampling exactly from the exponential distribution and
rounding the results to the closest floating point number.

However, if we represent the uniform deviates by u-rands, it is quite
easy to make the algorithm exact in the sense discussed in connection
with Table \ref{example} in Sec.~\ref{intro}.  A u-rand can be
represented in base $b$ as
\[
s(n + 0.d_0d_1\ldots d_{L-1} + \ldots),
\]
where $s = \pm1$, $n$ and $L$ are non-negative integers, $d_l$ are
digits in $[0,b)$, and the fraction is written in ordinary positional
notation.  Only the first $L$ digits of the fraction have been sampled
and the final ellipsis represents the digits which are not yet known,
i.e., it represents a sample from $b^{-L}U$.  Only a small number of
operations need to be implemented on u-rands to realize Algorithm V.
The assignment $x\leftarrow U$ in step V2 corresponds to $s\leftarrow1$,
$n\leftarrow0$, $L\leftarrow0$.  The operation $l + x$ in step V5 is
just $n\leftarrow l$.  This just leaves the comparisons between u-rands
in step V3.  In this case, we have $s=1$ and $n=0$ for both u-rands, so
only the fractions need to be compared.  The digits are compared
starting at position $0$ and if they are different, the comparison can
be made; if not, the next digits are compared.  During this process,
digits sampled uniformly from $[0,b)$ are added to the fractions and $L$
is incremented as necessary; typically, the comparisons can be made
examining only a few digits of each operand.

We shall need to add a few additional operations on u-rands to implement
the additional algorithms presented in this paper: negation
($x \leftarrow -x$), comparisons with a rational ($x < u/v$),
incrementing by one half ($x \leftarrow x + \frac12$); these are easily
accomplished (with the proviso that $b$ be even for the last operation).
We also need to be able to extract from a u-rand the value rounded to
the closest floating point number at some precision, sampling, if
necessary, additional digits.  This is straightforward if $b$ is a power
of two; all rounding modes can be supported (and the process provides a
flag indicating the direction of the rounding).  Finally, it is easy to
produce printed representations of the u-rand itself (e.g.,
``$+1.6\ldots$'') and of a correctly rounded fixed point representation
in base $b$ (for examples, see Table \ref{example}).  In this
connection, note that by using $b=10$ we are able to produce exactly
rounding decimal representations of normal deviates without any radix
conversions.

It's possible to make von Neumann's algorithm slightly more efficient by
using early rejection.
\algbegin Algorithm E (improved von Neumann).
Improved algorithm for sampling $E$ from a distribution with density
$e^{-x}$ for $x > 0$.
\algstep E1. [Initialize rejection count.] Set $l\leftarrow 0$.
\algstep E2. [Sample fraction.]  Set $x \leftarrow U$, where $U$
is a uniform deviate $U \in (0,1)$.
\algstep E3. [Early rejection.]  If $x > \frac12$, set
$l \leftarrow l + 1$ and go to step E2.
\algstep E4. [von Neumann's step V3.]  Sample uniform
deviates $U_1, U_2, \ldots$ and determine the maximum value $n \ge 0$
such that $x > U_1 > U_2 > \ldots > U_n$.
\algstep E5. [Test length of run.]  If $n$ is odd, set
$l \leftarrow l + 1$ and go to step E2.
\algstep E6. [Combine integer and fraction.]  Otherwise ($n$ is even),
set $y \leftarrow \frac12 l + x$.
\algstep E7. [Return result.] Set $E \leftarrow y$.%
\algend
The early rejection step results in lowering the mean number of uniform
deviates required to $e/(\sqrt e - 1) \approx 4.19$.

Von Neumann's algorithm can be adapted to generate a Bernoulli random
variable with probability $1/\sqrt e$, as follows.
\algbegin Algorithm H (a half exponential Bernoulli trial).
Generates a Bernoulli random value $H$ which is true with probability
$1/\sqrt e$.
\algstep H1. [Generate a run.] Sample uniform
deviates $U_1, U_2, \ldots$ and determine the maximum value $n \ge 0$
such that $\frac12 > U_1 > U_2 > \ldots > U_n$.
\algstep H2. [Test length of run.]  Set $H \leftarrow
(\text{$n$ is even}).$%
\algend
On average, the algorithm uses $\frac12 e/(\sqrt e - 1)$ (resp.~$\frac12
e$) uniform deviates if the result is false (resp.~true); the overall
weighted average is $\sqrt e$.  Both Algorithms E and H can be
implemented using u-rands and so can deliver exact results.  Algorithm H
will be used for sampling from the normal distribution.

\section{Sampling from the normal distribution}\label{sample}

Here I tackle the problem of sampling normal deviates using u-rands.
Although, I didn't realize it at the time, the method I developed is
closely related to the algorithm given by \citet[p.~41]{kahn56}; see
also \citet[\S26.8.6.a(4)]{abramowitz65}.  This is
\algbegin Algorithm K (Kahn).  Sample $N$ from a unit normal
distribution $\phi(x)$ using Kahn's method.
\algstep K1. [Sample absolute value of deviate $y$.] Set $y \leftarrow E$ 
where $E$ is an exponential deviate.
\algstep K2. [Adjust relative probability of $y$ by rejection.]  Sample
$z \leftarrow E$ and accept $y$ if $z > {\frac12(y-1)^2}$; otherwise go to
step K1.  (For a given $y$, the probability of acceptance
$\exp\bigl(-\frac12(y-1)^2\bigr)$.  Averaging over $y$, the probability
of acceptance is $\sqrt{\pi/(2e)} \approx 0.76$.)
\algstep K3. [Assign a sign.]  With equal probabilities, set
$x \leftarrow \pm y$.
\algstep K4. [Return result.] Set $N \leftarrow x$.%
\algend
A problematic step here is K2, which requires performing arithmetic on
$y$.  In order to avoid this, I found it necessary to sample separately
the integer and factional parts of $y$, leading to the following
skeleton of an algorithm.
\algbegin Algorithm N (normal sampling).  Sample $N$ from a unit normal
distribution $\phi(x)$ using a rejection method.
\algstep N1. [Sample integer part of deviate $k$.]  Select integer $k \ge 0$
with probability $\exp(-\frac12 k) (1 - 1/\sqrt e)$.
\algstep N2. [Adjust relative probability of $k$ by rejection.]
Accept $k$ with probability $\exp\bigl(-\frac12 k(k-1)\bigr)$; otherwise
go to step N1.
\algstep N3. [Sample fractional part of deviate $x$.]  Set
$x \leftarrow U$, where $U$ is a uniform deviate $U \in (0,1)$.
\algstep N4. [Adjust relative probability of $x$ by rejection.] Accept
$x$ with probability $\exp\bigl(-\frac12 x(2k+x)\bigr)$; otherwise go to
step N1.
\algstep N5. [Combine integer and fraction.]  Set $y \leftarrow k + x$.
\algstep N6. [Assign a sign.]  With probability $\frac12$, set
$y \leftarrow -y$.
\algstep N7. [Return result.] Set $N \leftarrow y$.%
\algend
The analysis of this algorithm is similar to that for Kahn's method.
After step N2, the relative probability density of $k$ is $\exp(-\frac12
k) \times {\exp\bigl(-\frac12 k(k-1)\bigr)} = \exp(-\frac12 k^2)$ for
$k \ge 0$; after step N4, the relative probability of $[k, x]$ is
$\exp(-\frac12 k^2) \times \exp\bigl(-\frac12 x(2k+x)\bigr) =
{\exp\bigl(-\frac12 (k + x)^2\bigr)}$ for $k \ge 0$ and $x \in (0, 1)$.
From this, it follows that the returned value of $x$ has a Gaussian
distribution, $\phi(x)$.  Step N2 always succeeds for $k = 0$ and $1$,
the two most common cases.  Overall, the probability that step N2
succeeds is $(1 - 1/\sqrt e) G \approx 0.690$ where $G
= \sum_{k=0}^\infty \exp(-\frac12 k^2) \approx 1.753$.  Similarly, step
N4 succeeds with probability $\sqrt{\pi/2} / G \approx 0.715$.  Thus,
step N1 is executed $\sqrt{2/\pi}/(1 - 1/\sqrt e) \approx 2.03$ times on
average.

Steps N1 and N2 can be expressed in terms of half exponential Bernoulli
trials $H$ with
\altalgbegin N (Steps N1 and N2 in terms of $H$.)
\algstep N1. [Test $H$ until failure.]  Generate a sequence of
Bernoulli deviates $H_1, H_2, \ldots$ and determine the largest $k \ge
0$ such that $H_1, H_2, \ldots, H_k$ are all $\mathrm{true}$.
\algstep N2. [Make $k(k-1)$ tests of $H$.]  Set
$k' \leftarrow k(k-1)$ and generate up to $k'$ Bernoulli deviates $H_1,
H_2, \ldots, H_{k'}$. Accept $k$ if $H_i$ is $\mathrm{true}$ for all
$i \in [1, k']$; otherwise go to step N1.
\altalgend

Of the remaining steps only step N4 presents a challenge.  This is
changed to
\altalgbegin N (Rewriting step N4.)
\algstep N4. [Break N4 into $k + 1$ steps.]  Perform up to $k+1$ Bernoulli
trials, $B_1, B_2, \ldots, B_{k+1}$, each with probability
${\exp\bigl(-x (2k + x)/(2k + 2)\bigr)}$.  Accept $x$ if $B_i$ is
$\mathrm{true}$ for all $i \in [1, k+1]$; otherwise go to step N1.
\altalgend
This transformation of N4 is motivated by the requirement in the proof
of von Neumann's method that $x \in [0,1]$.  Repeating the trial $k+1$
times means that the argument to the exponential in the original step N4
is divided by $k+1$; the maximum value of $x (2k + x)/(2k + 2)$ (as $x$
is varied) is $(2k + 1)/(2k + 2) < 1$.

In order to carry out a Bernoulli trial $B$, I generalize von Neumann's
procedure.
\algbegin Algorithm B (generalizing von Neumann's step V3).  A Bernoulli
trial with probability $\exp\bigl(-x (2k + x)/{(2k + 2)}\bigr)$.  Sample
two sets of uniform deviates $U_1, U_2, \ldots$ and $V_1,\allowbreak
V_2, \ldots$ and determine the maximum value $n \ge 0$ such that $x >
U_1 > U_2 > \ldots > U_n$ and $V_i < (2k + x)/(2k + 2)$ for all $i \in
[1,n]$.
\algstep B1. [Initialize loop.]  Set $y \leftarrow x$, $n \leftarrow 0$.
\algstep B2. [Generate and test next samples.]\\
(i) Sample $z \leftarrow U$; go to step B4, unless $z < y$.\\
(ii) Sample $r \leftarrow U$; go to step B4, unless $r < {(2k + x)}/(2k + 2)$.
\algstep B3. [Increment loop counter and repeat.] Set $y \leftarrow z$,
$n \leftarrow n + 1$; go to step B2.
\algstep B4. [Test length of runs.] Set $B \leftarrow
(\text{$n$ is even})$.%
\algend
Without step B2(ii), steps B1 to B3 are just step V3 of von Neumann's
algorithm.  Because of the additional test B2(ii), the probability that
the $n$th trip through the loop succeeds is $x^n/n! \times {\bigl((2k +
x)/(2k + 2)\bigr)^n}$.  The requirement that $n$ be even means that $B$
succeeds with probability
\ifeprint\begin{multline*}
1 - x \frac{2k + x}{2k + 2}
+ \frac{x^2}{2!} \biggl(\frac{2k + x}{2k + 2}\biggr)^2
- \frac{x^3}{3!} \biggl(\frac{2k + x}{2k + 2}\biggr)^3 + \ldots \\
= \exp\biggl(-x \frac{2k + x}{2k + 2}\biggr).
\end{multline*}\else
\[
1 - x \frac{2k + x}{2k + 2}
+ \frac{x^2}{2!} \biggl(\frac{2k + x}{2k + 2}\biggr)^2
- \frac{x^3}{3!} \biggl(\frac{2k + x}{2k + 2}\biggr)^3 + \ldots \\
= \exp\biggl(-x \frac{2k + x}{2k + 2}\biggr).
\]\fi

In order to avoid performing arithmetic on uniform deviates in step
B2(ii), we remark that as $x$ varies in $(0,1)$ the right side of the
inequality varies from $2k/(2k+2)$ to $(2k+1)/(2k+2)$.  Thus, regardless
of the values of $x$ and $r$, the test will succeed with probability
$2k/(2k+2)$ and fail with probability $1/(2k+2)$.  The remaining
probability, $1/(2k+2)$, is divided between success and failure
according to $r < x$.  Thus the test $r < (2k + x)/(2k + 2)$ can be
replaced with
\algbegin Algorithm T (the test in B2(ii)).  Perform test
$T = \bigl(r < (2k + x)/(2k + 2)\bigr)$ without doing arithmetic on real
numbers.
\algstep T1.  [Sample a selector $f$.] Set $f \leftarrow C(2k+2)$ where
$C(m)$ is $-1$ with probability $1/m$, $0$ with probability $1/m$, and
$1$ with probability $1-2/m$.
\algstep T2.  [Act on the value of $f$.]  If $f < 0$, set
$T \leftarrow \mathrm{false}$; else if $f > 0$, set
$T \leftarrow \mathrm{true}$; otherwise ($f = 0$), set $T \leftarrow {(r
< x)}$.%
\algend
Finally, $C(m)$ can be computed with
\algbegin Algorithm C (the 3-way selector).  The choice $C(m)$,
$(-1,0,1)$ with probabilities $(1/m, 1/m,\allowbreak {1-2/m})$,
implemented as the test $w < n/m$ where $w$ is a uniform deviate in
$(0,1)$ and $n = 1$ or $n = 2$.  For each successive digit $d$ of
$w$, substitute $w = (d+w')/b$ so that the test becomes $w' < n'/m$,
where $n' = bn - dm$, and exit as soon as the $n'$ is outside the range
$(0,m)$.
\algstep C1. [Set the numerators of the fractions.] Set
$n_1 \leftarrow 1$ and $n_2 \leftarrow 2$.
\algstep C2. [Sample the next digit of $w$, $d$.]  Sample
$d \leftarrow D$ where $D$ is a uniformly distributed integer in $[0,
b)$.
\algstep C3. [Multiply inequalities by $bm$.]  Set $n_1 \leftarrow bn_1 - dm$
and $n_2 \leftarrow bn_2 - dm$.
\algstep C4. [Test the new numerators.]  If $n_1 \ge m$,
set $C(m) \leftarrow -1$ and return; else if $n_2 \le 0$, set
$C(m) \leftarrow 1$ and return; else if $n_1 \le 0$ and $n_2 \ge m$,
set $C(m) \leftarrow 0$ and return; otherwise, go to step C2.%
\algend
Algorithm C shows how the comparison of a u-rand with a rational $x <
u/v$ can be implemented.  Step B2 can now be written as
\altalgbegin B (Step B2 incorporating Algorithm T.)
\algstep B2. [Generate and test next samples.]\\
(a) Sample $z \leftarrow U$; go to step B4, unless $z < y$.\\
(b) Set $f \leftarrow C(2k+2)$; if $f < 0$, go to step B4.\\
(c) If $f = 0$, sample $r \leftarrow U$ and go to step B4, unless
$r < x$.
\altalgend
The three steps here can be carried out in any order and I find that the
number of random digits needed can be reduced by reversing the order of
B2(a) and B2(b) whenever $k = 0$.

Now step N4 has been broken down into steps that can all be carried out
in terms of u-rands.  In the final step, N7, the normal deviate can be
returned either as a u-rand or an exactly rounded floating point number.

\section{Sampling from the discrete normal distribution}\label{discrete}

In some applications, we wish to sample integers, $i$, from the discrete
normal distribution,
\[
\phi(i\mid\mu,\sigma) \propto
\exp\biggl[-\frac12\biggl(\frac{i - \mu}{\sigma}\biggr)^2\biggr],
\]
which is characterized by parameters $\mu$ and $\sigma$.  In the limit
$\sigma\gg1$, the mean and variance of this distribution is well
approximated by $\mu$ and $\sigma^2$.  Considering the class of integer
distributions, this distribution maximizes the entropy for a given mean
and variance \citep{kemp97}.  Because Algorithm N is a simple rejection
scheme, it is rather easy to adapt it to sample from the discrete
distribution as follows:
\algbegin Algorithm D (discrete normal sampling).  Sample $D$ from a
discrete normal distribution $\phi(i\mid\mu,\sigma)$ using a rejection
method.
\algstep D1. [Same as step N1.]  Select integer $k \ge 0$
with probability $\exp(-\frac12 k) (1 - 1/\sqrt e)$.
\algstep D2. [Same as step N2.]
Accept $k$ with probability $\exp\bigl(-\frac12 k(k-1)\bigr)$; otherwise
go to step D1.
\algstep D3. [Assign a sign.] With equal probabilities,
set $s \leftarrow \pm 1$.
\algstep D4. [Sample fractional part of deviate $x$.]  Set
$x \leftarrow x_0 + j/\sigma$, where $x_0 = \bigl(i_0 - (\sigma k +
s\mu)\bigr) / \sigma$, $i_0 = \ceil{\sigma k + s\mu}$, and $j$ is a
random integer uniformly sampled from $\bigl[0,\ceil\sigma\bigr)$.
\algstep D5. [Ensure that $x$ is in the allowed range.] 
If $x \ge 1$, go to step D1.  (This cannot happen if $\sigma$ is
an integer.)
\algstep D6. [Avoid double counting 0.]  If $k = 0$, $x = 0$,
and $s < 0$, go to step D1.  (This cannot happen unless $\mu$ is
an integer.)
\algstep D7. [Same as step N4.] Accept $x$ with probability
$\exp\bigl(-\frac12 x(2k+x)\bigr)$; otherwise go to
step D1.
\algstep D8. [Combine parts of the integer deviate.]  Set
$i \leftarrow s(i_0 + j)$.
\algstep D9. [Return result.] Set $D \leftarrow i$.%
\algend
This is nearly the same as Algorithm N, except that step N3 has been
replaced by steps D4--D6.  These steps can be understood by matching
$\phi(x)$ and $\phi(i\mid\mu,\sigma)$, identifying $s(k+x) =
\bigl({s(i_0+j)}-\mu\bigr)/\sigma$, and determining $i_0$ and $j$ such that
$x\in[0,1)$.  When $j=0$, $x$ takes on the value $x_0$ and $i_0$ is that
integer which minimizes $x_0$ while maintaining the condition $x_0\ge0$.
Similarly the requirement that $x<1$, imposes the condition
$j<\ceil\sigma$.  If $\sigma$ is not an integer, then for some
values of $k$ and $s$, there are only $\floor\sigma$ allowed
values of $j$, so, in step D5, we enforce the condition $x\in[0,1)$.
Finally, in step D6, we avoid double counting the origin of the normal
distribution by additionally requiring that $x\in(0,1)$ if $k = 0$ and
$s < 0$.

Algorithm D is a straightforward modification of Algorithm N and all the
steps can be carried out exactly if the parameters $\mu$ and $\sigma$
are rational.  One additional function needs to be added to the
machinery to handle u-rands, namely a comparison with a rational (this
is needed in implementing Algorithm B because now $x$ is a rational
number).  See Algorithm C for how this can be implemented.

Because the probability that step D7 (i.e., step N4) succeeds is
$0.715$, then for $\sigma$ large (and $b = 2$), this method requires at
least $(1/0.715)\log_2\sigma$ bits of randomness on average.  We would
like to reduce the multiplier of $\log_2\sigma$ from $1/0.715$ to $1$ to
match the perfect scaling of Algorithm N.  We can achieve this goal by
adapting the algorithm given by \citet{lumbroso13} for sampling an
integer in $[0,m)$.  Sufficient of his algorithm is carried out to allow
the result to be returned as range of size $b^l$.  For example, if $b =
2$, sampling from $[0,9)$ returns the ranges $\{[0,8),\allowbreak
[0,2),\allowbreak [2,6),\allowbreak [6,8),\allowbreak [8,9)\}$ with
probabilities $\{\frac{32}{63},\allowbreak \frac2{21},\allowbreak
\frac4{21},\allowbreak \frac2{21},\allowbreak \frac19\}$.
Thereafter the range can be narrowed, if necessary, by factors of $b$ to
allow the inequality in step D7 to be evaluated.  This method of
sampling integers generalizes the concepts of a u-rand to apply to
discrete sampling and, if Algorithm D is implemented using this
technique, it exhibits perfect scaling in the limit of large~$\sigma$.

As an aside, this technique of partially sampling a discrete uniform
distribution allows optimization of sampling from the Bernoulli
distribution with rational probability $p = u/v$.  Conventionally, this
is implemented by testing $u<j$ with $j$ sampled uniformly in $[0,v)$
which for $v$ large requires at least $\log_2v$ random bits.  However if
$j$ is partially sampled, the mean cost is a constant in the limit of
large $v$.  Similar performance is obtained with u-rands, sampling
$x \leftarrow U$ and testing $x < u/v$ (and, with u-rands, this test is
exact).

\section{Implementation}\label{implement}

An implementation of Algorithm N in C++11 is available in the library
ExRandom.  This allows the user to select the base $b$ and to access the
random deviate as a u-rand or as a floating point number.  The algorithm
has also been wrapped into a C++11 ``random number
distribution'' \citep[\S40.7.3]{stroustrup13}, \verb|unit_normal_distribution|,
which can be used as a replacement for the
standard \verb|normal_distribution| (with zero mean and unit variance).
Several test programs are also provided, one of which performs the
$\chi^2$ test on the output of Algorithm N, which is an essential step
in validating its correctness.  This test passes with $10^{10}$ samples
and 50 bins of equal width in the interval $[-4,4]$ when the Mersenne
Twister random number generator \citep{matsumoto98}, \verb|mt19937|, is
used as the source of random digits.  (On the other hand, the test fails
badly with $10^9$ samples using the linear congruential
generator, \verb|minstd_rand0|.)  Algorithm N retains no state from one
invocation to the next.  So it is not necessary to verify the
independence of the normal deviates (any lack of independence would be
due to defects in the underlying random generator).

In addition, Algorithm N has also been incorporated in \citet{MPFR},
version 3.2.0, a library for arbitrary precision
arithmetic \citep{fousse07} as the function \verb|mpfr_nrandom|.  MPFR,
version 3.1.0, already provided a function \verb|mpfr_grandom| for
sampling normal deviates based on the polar method \citep{box58}.

\begin{table}[tb]
\tbl{\label{timing}Times (in $\mathrm{\mu s}$) for sampling from the normal
distribution.  The quantity $p$ is the number of bits in the fraction of
the rounded floating point samples.  Columns A and D use the polar
method, while columns B and C use Algorithm N.  Columns A and B time the
C++11 random number distributions {\tt normal\_distribution} and {\tt
unit\_normal\_distribution} delivering results in the form of IEEE
floating point numbers.  Columns C and D time the routines {\tt
mpfr\_nrandom} and {\tt mpfr\_grandom} which produce MPFR floating point
numbers.  Columns A and B are the results of averaging over $50$ million
samples; the entries in Columns C and D are each the result of averaging
over about $10$ seconds.}{
\begin{tabular}{ccccc}
\hline\hline\noalign{\smallskip}
type & \multicolumn{2}{c}{IEEE} & \multicolumn{2}{c}{MPFR}\\
method & polar & \multicolumn{2}{c}{Algorithm N} & polar
\\\hline\noalign{\smallskip}
$p$ & A & B & C & D
\\\hline\noalign{\smallskip}
$24$     & $0.034$ & $0.30$ & $0.59$ & $2.3$    \\
$32$     &         &        & $0.64$ & $2.4$    \\
$53$     & $0.054$ & $0.31$ & $0.64$ & $2.6$    \\
$64$     & $0.057$ & $0.37$ & $0.64$ & $2.8$    \\
$128$    &         &        & $0.65$ & $3.8$    \\
$256$    &         &        & $0.68$ & $6.2$    \\
$2^{10}$ &         &        & $0.86$ & $20$     \\
$2^{12}$ &         &        & $1.6$  & $130$    \\
$2^{14}$ &         &        & $4.3$  & $1300$   \\
$2^{16}$ &         &        & $15$   & $13000$  \\
$2^{18}$ &         &        & $59$   & $120000$ \\
$2^{20}$ &         &        & $240$  & $910000$ \\
\hline\hline
\end{tabular}}
\end{table}
Table \ref{timing} shows some comparative timings for producing normal
deviates with a precision of $p$ bits.  The tests were run on a Fedora
Linux system with a $3.2\,\mathrm{GHz}$ Intel processor using the g++
compiler version 4.8.2.  In all cases, the Mersenne Twister algorithm
was used to generate the random digits and the implementations of
Algorithm N timed here use $b = 2^{32}$ to match the output of this
generator.  Comparing columns A and B, we see that Algorithm N is an
order of magnitude slower than the polar method at producing double
precision results.  On the other hand (comparing columns C and D),
Algorithm N is dramatically faster than \verb|mpfr_grandom| at producing
arbitrary precision results in the MPFR format.  As expected, the
scaling of the time for Algorithm N in column C is offset linear,
approximately $(1 + 240\,p/2^{20})\, \mathrm{\mu s}$.

It is instructive to compare the two MFPR routines \verb|mpfr_grandom|
and \verb|mpfr_nrandom|.  The former provides a good illustration of how
a conventional method for sampling random deviates can be implemented
with guarantees on the accuracy; the working precision needs to be
progressively increased; and, of course, heavy use is made of the
formidable infrastructure provided by MPFR for carrying out arbitrary
precision arithmetic.  The final result entails computing a logarithm
and extracting a square root which incur a reasonably heavy penalty as
the precision is increased (the time increases roughly as $p^{1.6}$).
On the other hand, \verb|mpfr_nrandom| relies on MPFR only to provide
the data type to hold the result and the penalty for high precision is
minimal.

Table \ref{timing} times the production of normal deviates in a standard
computational environment in which uniform pseudo random numbers can be
produced rather rapidly.  In some security applications, it may be
necessary to use a slow hardware random number generator.  In this case,
Algorithm N can be used with $b = 2$ to conserve random bits.  Let $B$
be the number of bits consumed by Algorithm N and $L$ be the number of
bits in the fraction of resulting u-rand.  Empirically, I find that
$\ave B \approx 30.000$ and $\ave L \approx 1.556$.  The distribution of
$L$ decays with an $e$-folding constant of $1/\log 2$, while that of $B$
decays more slowly with a $e$-folding constant of about $29.9$ bits.  To
put these results in perspective, if $10^{30}$ normal deviates were
generated, then the largest result would be $\abs x \sim 12$, the
longest fraction would have $L \sim 100$ and at most $B \sim 2000$ bits
would be needed to generate a single normal deviate.

The quantity $C = \ave B - \ave L \approx 28.444$ represents the
``cost'' of producing random deviates.  Producing rounded fixed point
normal deviates with $p$ bits in the fraction requires $C + p + 1$
random bits on average; this formula applies for large $p$ (but $p \ge
10$ suffices in practice).  The $1$ here accounts for the additional bit
needed for rounding the result (and the rounding operation, in turn,
provides an extra bit of information, namely whether the true deviate is
larger or smaller than the rounded result).  Producing rounded floating
point normal deviates with precision $p$ requires $C - Q + p + 1$ bits
on average (the $1$ again accounts for the need for a rounding bit);
here $Q = \ave{\floor{\log_2\abs x}} + 1 \approx -0.417$ is the mean
floating point exponent for normal deviates.  Thus producing IEEE double
precision floating point numbers ($p = 53$) requires about $82.861$ bits
per rounded deviate, on average.

When comparing Algorithm N with algorithms for {\it other}
distributions, we use the toll of the algorithm defined as $T = C -
H \approx 26.397$ where $H = \log_2\sqrt{2\pi e} \approx 2.047$ is the
entropy of the normal distribution and the base-$2$ logarithm is used so
that $H$ is measured in bits.  A {\it perfect} sampling algorithm would
have $T=0$; so the toll is a measure of how many random bits are
potentially ``wasted'' by the algorithm.  (Note that the entropy of the
discrete distribution obtained by rounding normal deviates to the
closest multiple of $2^{-p}$ is $H + p$, for $p$ large.)

\begin{figure}[tp]
\centerline{\includegraphics[scale=0.75]{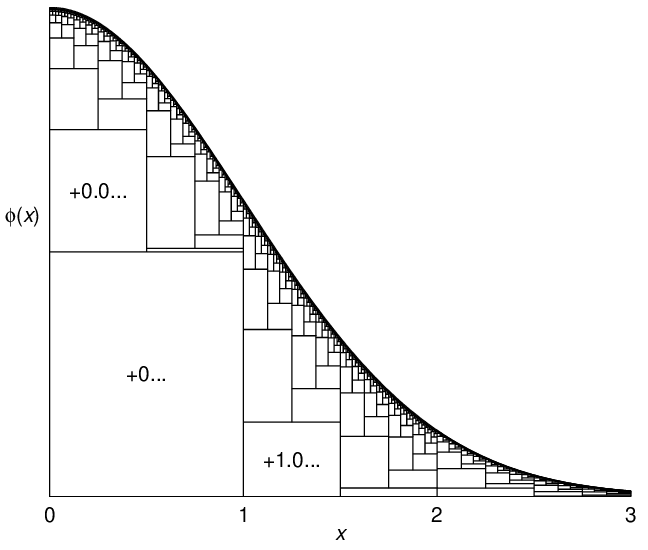}}
\caption{Algorithm N's decomposition of the normal distribution into a set of
uniform distributions with $b = 2$ (shown for $x > 0$).  For example,
the frequency with which $+0.0\ldots$ is returned is equal to the
relative area of the rectangle spanning $x \in (0,\frac12)$ which is
$\frac18\sqrt{2/\pi} \approx 10\%$.  This frequencies used in this
figure are averaged over $10^{10}$ samples; the minimum range of the
uniform distributions shown is $2^{-8}$.}
\label{normalhist}
\end{figure}
A histogram of the u-rands that Algorithm N produces can be displayed in
a way that illustrates how they add up to the normal distribution as
shown in Fig.~\ref{normalhist}.  The area assigned to each u-rand is
proportional to its frequency while its base covers its range (here $b =
2$).  This figure is reminiscent of illustrations of the ziggurat
method \citep{marsaglia00}, which provides a fast way of sampling normal
deviates by {\it approximating} the normal distribution by a {\it
static} set of rectangles enabling it to return a uniform deviate in
{\it most} cases.  In contrast, Algorithm N {\it dynamically} generates
a set rectangles which cover the normal distribution {\it exactly},
allowing it to return a uniform deviate in {\it all} cases.

The ExRandom library includes implementations of Algorithms V and E for
sampling from the unit exponential distribution with interfaces that
parallel those for the unit normal distribution.  The C++ random number
distribution \verb|unit_exponential_distribution| uses Algorithm V
because it is slightly faster than Algorithm E with $b=2^{32}$.  This
produces double precision deviates in $0.09\,\mathrm{\mu s}$ under the
same conditions as in Table~\ref{timing}.  (Algorithm V has also been
added to MPFR, version 3.2.0, as \verb|mpfr_erandom|.)  With $b=2$,
Algorithm E results in $\ave B \approx 7.232$ and $\ave L \approx
1.743$, so the cost is $C \approx 5.489$.  For the exponential
distribution, we have $H = \log_2 e \approx 1.443$ and $Q \approx
-0.333$.  Thus the toll is $T = C-H \approx 4.047$ (considerably better
than Algorithm N) and producing IEEE double precision numbers requires
$C-Q+54 \approx 59.822$ bits on average.  In contrast, von Neumann's
original method, Algorithm V, has a cost $C \approx 7.262$ and a toll
$T \approx 5.819$; i.e., it is less efficient than Algorithm E by about
$1.772$ bits.



Finally, the ExRandom includes implementations of Algorithm D for
sampling from the discrete normal distribution.  The parameters $\mu$
and $\sigma$ are given as the ratio of two 32-bit integers.  However
some internal calculations use 64-bit integers in an effort to avoid
overflow.  The constructors for the class implementing this algorithm
throw an exception if the parameters are such that overflow is possible.
The most stringent of the checks is that $b k_{\mathrm{max}}
\sigma_{\mathrm{num}}$ fits in a 64-bit word where $b$ is the base,
$k_{\mathrm{max}} = 50$ is how many standard deviations onto the tail of
the normal distribution we want to be able to sample, and
$\sigma_{\mathrm{num}}$ is the numerator of $\sigma$ when it and $\mu$
are expressed with a common denominator.

The C++11 random number distribution implementing Algorithm D uses $b =
2^{16}$.  The time to generate discrete normal deviates depends weakly
on $\sigma$ over the range $[1.6, 1.6\times10^6]$ varying between
$0.4\,\mathrm{\mu s}$ and $0.5\,\mathrm{\mu s}$ under the same
conditions as in Table~\ref{timing}.  With $b = 2$, the toll, defined
now merely as the difference between the mean number of bits to obtain a
discrete normal deviate and the entropy of the distribution (in bits),
is, in the limit of large $\sigma$, approximately a periodic function of
$\log_2\sigma$ with period $1$, attaining its minimum value of about
$27.9$ when $\sigma$ is a power of two and its maximum value of $31.9$
when $\sigma$ slightly exceeds a power of two.

An important potential use for Algorithm D is in cryptography, where
exact sampling is often required.  One such application is the
``learning with errors'' (LWE) problem \citep{regev09}, which depends on
the difficulty of solving a system of over-determined linear equations
over the field of integers when the equations have been perturbed by
noise sampled from a discrete normal distribution.  The security of the
cryptographic methods based on the LWE problem depends, in part, on
being able to sample discrete normal deviates accurately.  In some such
applications, there is also the requirement that the sampling algorithm
run on devices without hardware support for floating point operations.
Methods for sampling from the discrete normal distribution have recently
been reviewed by \citet{dwarakanath14}.  However, one of starting points
of this paper that ``sampling algorithms require either high precision
floating point arithmetic or very large precomputed tables'' is directly
contradicted by Algorithm D; it uses no floating point arithmetic and
requires no stored constants.  The potential drawbacks of Algorithm D in
this context are: (1)~The toll is large compared to implementations of
the method discrete distribution generating (DDG) trees
of \citet{knuth76} for which the toll is $2$; but that method is
impractical for large $\sigma$ because it involves storing large
precomputed tables of probabilities.  (2)~The time to generate a deviate
is variable (potentially leaking information to an attacker); this can
be mitigated by generating the deviates in batches of a thousand, say.
(3)~The time and memory requirements of the algorithm are unbounded;
but, with a slight reduction in accuracy, it is easy to put bounds on
these.  For example, if the number of digits in the fraction of u-rands
is limited to $\log N/\log 2$ and if the number of random digits allowed
for a single deviate to limited to $30\log N$, then the limits are hit
about once every $N$ invocations.  Even if $N$ is large, say $10^{30}$,
the resulting limits are modest; in the the rare cases when the limits
are hit, an integer uniformly sampled in
$[\floor{\mu-\sigma}, \ceil{\mu+\sigma}]$ can be returned.

\section{Conclusions}

I have presented an algorithm for sampling normal deviates with an
astonishing combination of properties: it is exact, it can be
implemented in a few dozen lines of code using only simple integer
operations, and it is fast.  The definition of ``exact'' is rigorous and
this property depends only on the availability of a source of uniform
random numbers.  Although the algorithm is an order of magnitude slower
at producing double precision results compared to conventional (less
accurate) methods, this is partly due to the hardware support provided
for floating point operations on modern computers.  If there is no such
support, as is the case for higher precision float point formats,
Algorithm N becomes competitive; indeed in the limit of high precision,
the only cost is that to produce and copy the random bits into the
result.

Algorithm N probably won't be useful in most routine Monte Carlo
simulations where the accuracy of conventional double precision methods
suffices.  However, in some specialized applications, the need for
accuracy is paramount.  In particular, the discrete version of the
algorithm, Algorithm D, has direct applications to some areas of
cryptography and here the fact that the algorithm entails only integer
instructions would also allow it to be implemented for embedded devices.

Internally, these algorithms represent real random deviates in terms of
u-rands and the floating point result is extracted from these at the end
of the algorithm.  This extraction process takes time, consumes memory,
and involves a round off error.  So, it might be advantageous to leave
the result as a u-rand; this occupies $O(1)$ storage and is still exact.
Furthermore, certain operations can be performed on u-rands at $O(1)$
cost.  For example, when implementing Algorithm K, the exponential
deviates $y$ and $z$ should be sampled as u-rands using Algorithm E.
The comparison in step K2 requires arbitrary precision arithmetic;
however, it can be completed with the addition of only a few extra
digits to $y$ and $z$, on average.  This means that it shares with
Algorithm N the perfect scaling of cost with precision.  As another
example, consider the operation $y \leftarrow x + \frac13$ where $x$ is
the (base~$2$) u-rand $+0.0\ldots$, i.e., a random deviate in the range
$[0,\frac12]$.  Carrying this out with floating point arithmetic entails
three rounding errors (for $x$, $\frac13$, and the sum) and involves
three $O(p)$ operations.  Alternatively, we could repeatedly sample
$y\leftarrow U$ until the conditions $\frac13 < y < \frac56$ are
satisfied, yielding an exact result in $O(1)$ operations.  Thus, it
would be of interest to explore the algebra of operations on u-rands.
The resulting ``lazy evaluation'' framework would, in principle, require
less storage, be faster, and be exact.

Algorithms E and N constitute a new class of algorithms for sampling
from continuous distributions offering the advantages of exactness and
perfect scaling.  Algorithm N builds on von Neumann's work adding two
new techniques: (1) breaking step N4 into $k+1$ tests, to reduce the
argument of the exponential; and (2) adding a second set of tests, in
step B2(ii), to compute a more complex exponential probability.
Presumably similar algorithms can be found for other distributions
although, as yet, there is no systematic approach to finding such
algorithms.  Related work by \citet{flajolet11} discusses several
interesting methods for sampling discrete distributions and considers
ways in which they can be combined.  It's probable that some of their
techniques will be useful in finding algorithms for sampling from other
continuous distributions; they might also lead to improvements to
Algorithm N for normal deviates.

\def\ACKNOWLEDGMENTS{I would like to thank Damien Stehl\'e for
pointing out the applications to cryptography and for drawing my
attention to Kahn's algorithm for sampling from the normal
distribution.}

\ifeprint
\section*{Acknowledgment}
\ACKNOWLEDGMENTS
\else
\begin{acks}
\ACKNOWLEDGMENTS
\end{acks}
\fi

\bibliography{rand}

\end{document}